\documentclass[runningheads]{llncs}
\usepackage[T1]{fontenc}
\usepackage{graphicx}
\usepackage{subfigure}
\usepackage{xcolor}
\begin{document}
\title{Deep Learning for Glioblastoma Morpho-pathological Features Identification: A BraTS-Pathology Challenge Solution\thanks{Corresponding author: Ying Weng.}}
\titlerunning{Deep Learning for Glioblastoma Morpho-pathological Features Identification}
%
\author{Juexin Zhang\orcidID{0000-0001-9086-7342} \and
Ying Weng\orcidID{0000-0003-4338-713X} \and
Ke Chen\orcidID{0000-0002-2046-0034}}
\authorrunning{J. Zhang et al.}
%
\institute{University of Nottingham Ningbo China, Ningbo 315100, China
\email{\{juexin.zhang,ying.weng,ke.chen2\}@nottingham.edu.cn}\\}

\maketitle              
\begin{abstract}
Glioblastoma, a highly aggressive brain tumor with diverse molecular and pathological features, poses a diagnostic challenge due to its heterogeneity. Accurate diagnosis and assessment of this heterogeneity are essential for choosing the right treatment and improving patient outcomes. Traditional methods rely on identifying specific features in tissue samples, but deep learning offers a promising approach for improved glioblastoma diagnosis. In this paper, we present our approach to the BraTS-Path Challenge 2024. We leverage a pre-trained model and fine-tune it on the BraTS-Path training dataset. Our model demonstrates poor performance on the challenging BraTS-Path validation set, as rigorously assessed by the Synapse online platform. The model achieves an accuracy of 0.392229, a recall of 0.392229, and a F1-score of 0.392229, indicating a consistent ability to correctly identify instances under the target condition. Notably, our model exhibits perfect specificity of 0.898704, showing an exceptional capacity to correctly classify negative cases. Moreover, a Matthews Correlation Coefficient (MCC) of 0.255267 is calculated, to signify a limited positive correlation between predicted and actual values and highlight our model's overall predictive power. Our solution also achieves the second place during the testing phase.

\end{abstract}
\section{Introduction}

Brain tumors are growths of abnormal cells within the brain. These can be either benign (non-cancerous) or malignant (cancerous). Glioblastoma, the most common malignant brain tumor, presents a significant challenge due to its aggressive nature. These tumors rapidly proliferate, infiltrate healthy brain tissue, and have a poor prognosis. The presence of distinct subtypes within a single glioblastoma (GBM), a phenomenon known as heterogeneity, significantly complicates treatment. This internal diversity fuels their aggressiveness and contributes to short survival rates.

Brain tumor pathology acts as the cornerstone of diagnosis, wielding microscopic analysis of tissue samples to identify the specific type of tumor, ultimately guiding crucial treatment decisions. Digital pathology is the evolution of diagnosing diseases by examining tissues under a microscope. Instead of traditional glass slides, it utilizes high-resolution digital scans of these tissue samples. Digital pathology is a rapidly evolving field which is transforming brain tumor diagnosis. By creating high-resolution digital scans of tissue samples, it empowers artificial intelligence to analyze the images for subtle features, potentially leading to faster and more accurate diagnoses.

In a bid to create AI tools that can identify distinct histological regions within brain tumors, the BraTS-Pathology Challenge \cite{bakas2024bratspath} has been launched. In this paper, we introduce a transfer learning method that utilizes a ResNet-18 \cite{he2016deep} pretrained on ImageNet \cite{deng2009imagenet} to achieve accurate classification of specific histological areas of interest. The remainder of the paper is organized as follows: our detailed methodologies are described in Section \ref{methods}. Section \ref{results} presents the experimental results of our proposed model, and the paper is concluded in Section \ref{conclusion}.

\section{Method}\label{methods}
\subsection{Network Architecture}

\begin{figure}[h!]
    \centering
    \includegraphics[width = 0.5\textwidth]{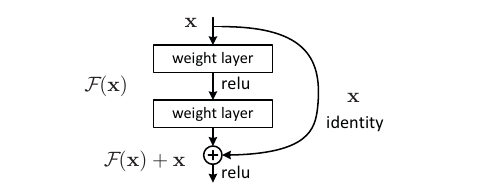}
    \caption{Residual block.}
    \label{residual}
\end{figure}

\begin{figure}
    \centering
    \includegraphics[width = \textwidth]{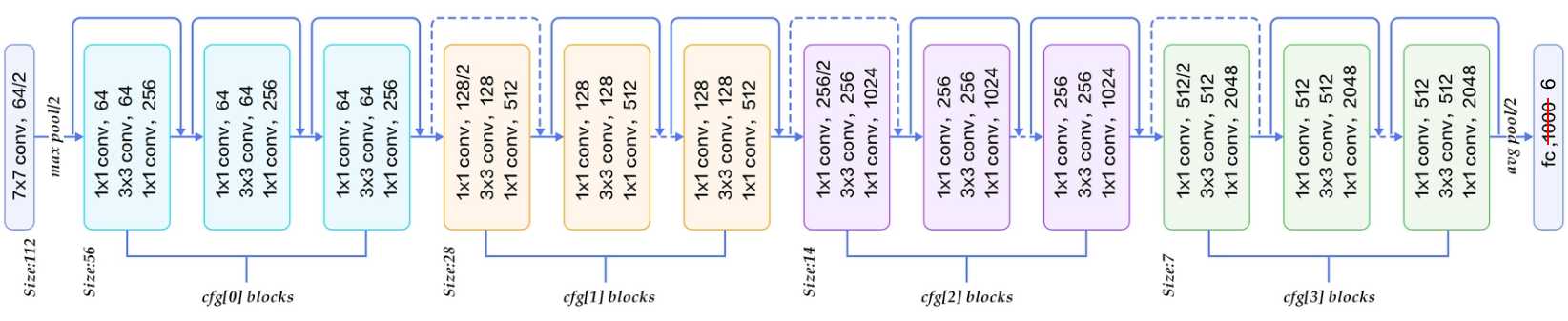}
    \caption{Architecture of our model. We replace the fully connected layer of ResNet-18 to output a 6-dim vector.}
    \label{resnet}
\end{figure}

A pre-trained ResNet-18 architecture, as proposed by He et al. \cite{he2016deep}, is served as the foundation for our model. To initiate the process, input data is downscaled to a 256x256 resolution through a 2D convolutional layer equipped with a 7x7 kernel. The core of our model consists of multiple residual blocks, a hallmark of ResNet design. Each of these blocks incorporates shortcut connections that enable identity mapping, facilitating the direct addition of outputs from stacked layers to the block's output (Fig. \ref{residual}). Within each residual block, two 3x3 convolutional layers are employed, followed by a batch normalization layer to stabilize training and accelerate convergence. Notably, the number of feature channels is doubled after every two consecutive residual blocks, while the spatial dimensions of the feature maps are halved. Subsequent to the residual blocks, a global average pooling layer is applied to reduce the feature maps to a single 1x1 vector. Finally, a fully connected layer projects these features into a 6-dimensional space, corresponding to the probability distribution of each class. A visual representation of the model architecture is provided in Fig. \ref{resnet}.

\section{Experiments}\label{results} 
\subsection{Dataset}
In the BraTS-Path challenge \cite{bakas2024bratspath}, publicly available H\&E-stained FFPE digitized tissue sections from The Cancer Imaging Archive's TCGA-GBM and TCGA-LGG collections are used. The dataset includes a retrospective, multi-institutional cohort of patients with diffuse gliomas, providing a clinically rich foundation for the challenge. 

To align with the latest 2021 World Health Organization (WHO) classification of central nervous system (CNS) tumors, the collections are reclassified. This involves identifying two key groups: glioblastoma multiforme (GBM) cases lacking the IDH mutation (IDH-wildtype) within CNS WHO grade 4, and low-grade astrocytomas from The Cancer Genome Atlas - Lower Grade Glioma (TCGA-LGG) that harbors specific molecular markers indicative of a more aggressive GBM. Including these reclassified astrocytomas ensures our algorithms function effectively for all clinical GBMs as defined by the WHO. The reclassified datasets exclude cases from the original TCGA-GBM collection whose molecular profiles do not align with the current WHO definition of GBM. Analyses focus on a single H\&E-stained tissue section per case to minimize artifacts associated with frozen tissue sections.

Expert neuropathologists have meticulously annotated tissue sections diagnosed as glioblastoma. These sections are divided into uniform patches, categorized either by specific tissue types or labeled as 'background' for areas lacking distinct features. Each patch is considered an individual case for in-depth analysis, enabling a fine-grained exploration of the glioblastoma's diverse characteristics. The annotated histologic areas of interest are as follows:
\begin{itemize}
\item presence of cellular tumor (CT).
\item pseudopalisading necrosis (PN).
\item areas abundant in microvascular proliferation (MP).
\item geographic necrosis (NC).
\item infiltration into the cortex (IC).
\item penetration into white matter (WM).
\end{itemize}

\begin{figure}[h!]
    \centering
    \includegraphics[width=0.6\textwidth]{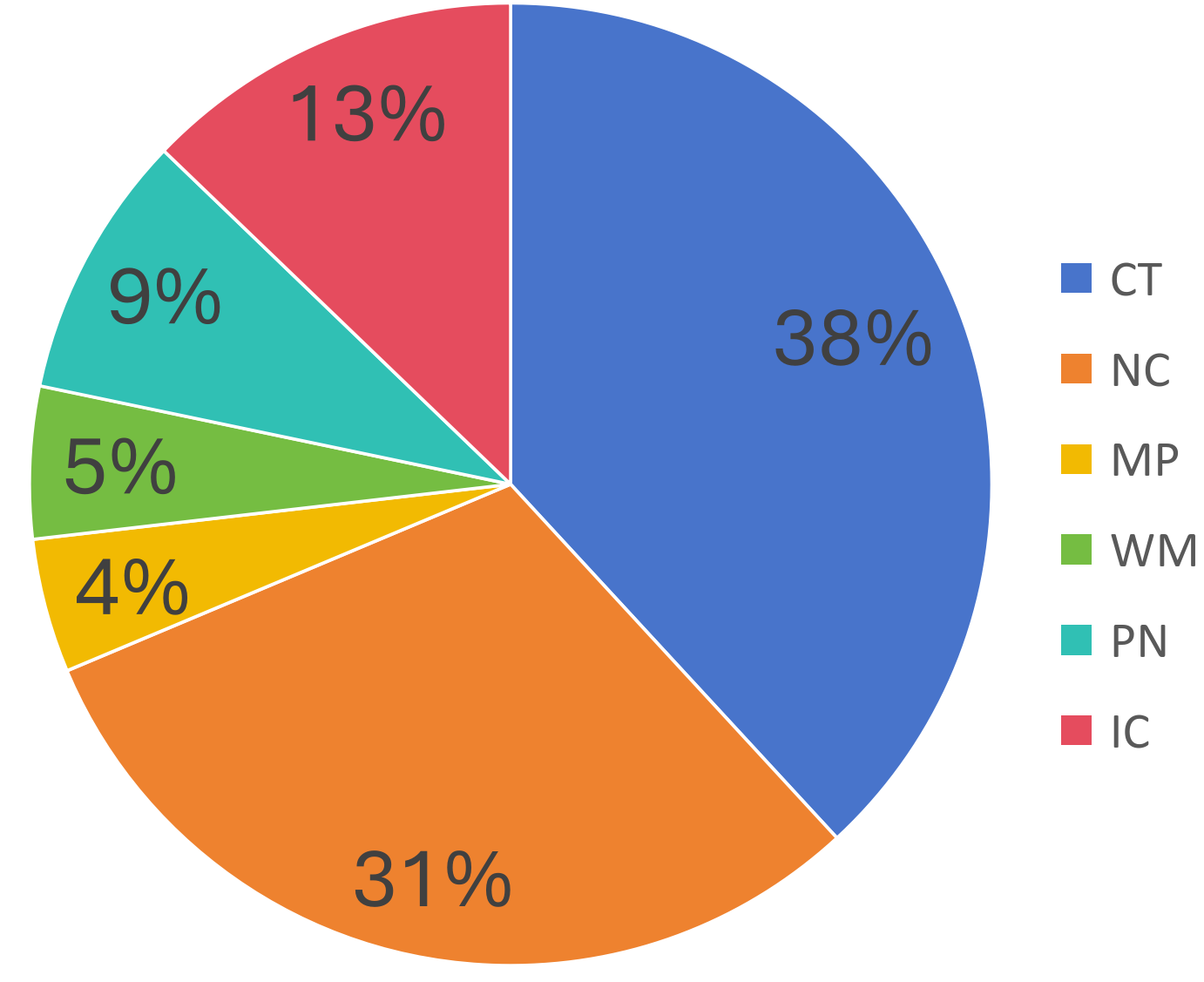}
    \caption{Percentage of each class.}
    \label{pie}
\end{figure}

The class distribution is determined and visualized in Fig. \ref{pie}. The dataset exhibits class imbalance, with CT and NC being the dominant classes.

\begin{figure}[h!]
\centering
\subfigure[CT]{
\includegraphics[width=0.25\textwidth]{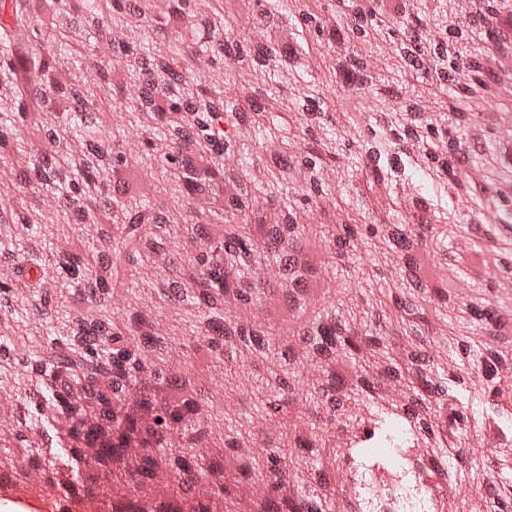}}
\hspace{0.05\textwidth}
\subfigure[PN]{
\includegraphics[width=0.25\textwidth]{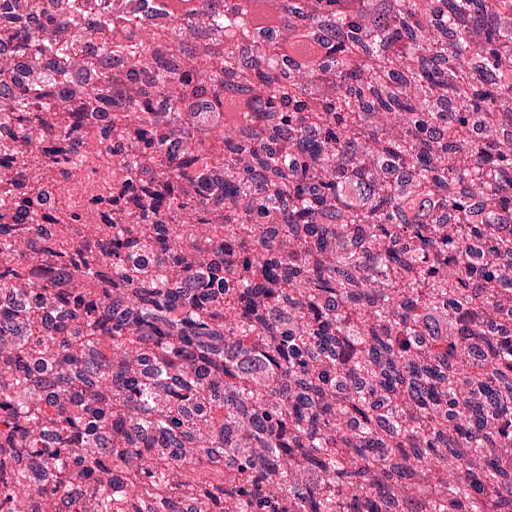}}
\hspace{0.05\textwidth}
\subfigure[MP]{
\includegraphics[width=0.25\textwidth]{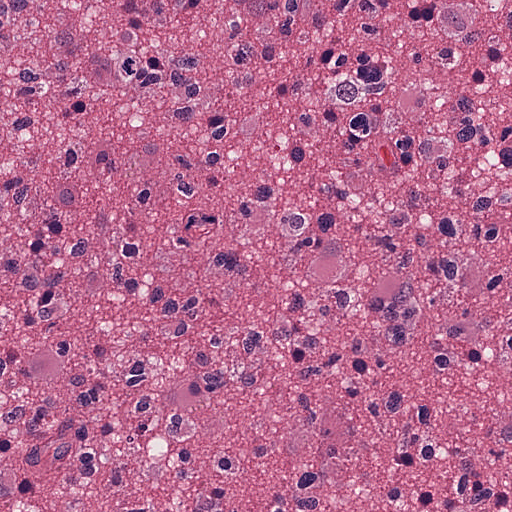}}

\subfigure[NC]{
\includegraphics[width=0.25\textwidth]{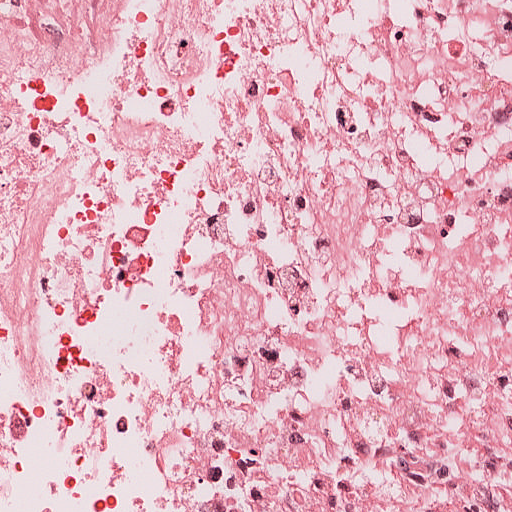}}
\hspace{0.05\textwidth}
\subfigure[IC]{
\includegraphics[width=0.25\textwidth]{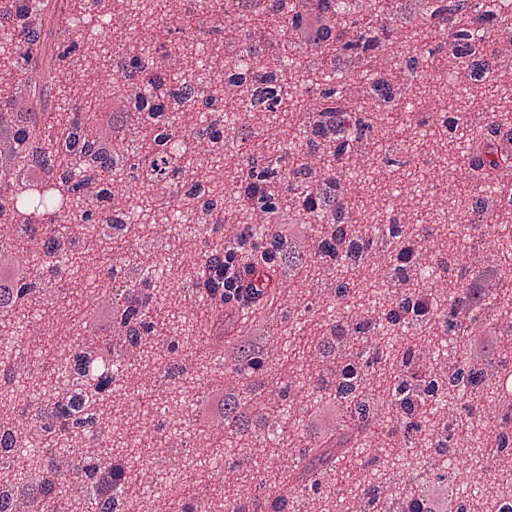}}
\hspace{0.05\textwidth}
\subfigure[WM]{
\includegraphics[width=0.25\textwidth]{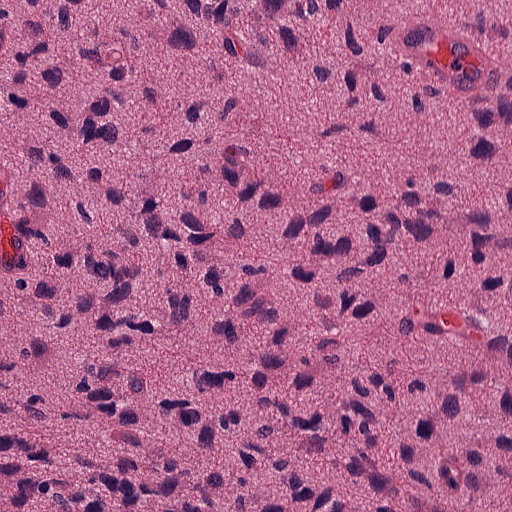}}
\caption{ The annotated histologic areas of interest.}
\label{fig2}
\end{figure}

\subsection{Data pre-processing}
Image preprocessing commenced by reading image files uses the OpenCV library. Given OpenCV's native BGR color format, a transformation to the RGB color space is subsequently performed to align with the model's input expectations. To ensure numerical stability and compatibility with subsequent processing steps, the pixel intensity values are scaled from the original 8-bit integer range of [0, 255] to a floating-point representation within the interval [0, 1]. Finally, to mitigate the impact of varying pixel intensity distributions across images and enhance model generalization, dataset normalization is applied. This process centers the pixel values at zero mean and scales them to unit standard deviation.

\subsection{Evaluation Metrics}
To comprehensively assess the performance of our model, a suite of evaluation metrics is employed. These metrics provide insights into various aspects of the model's predictive capabilities. Specifically, accuracy, recall, F1-score, specificity, and the Matthews Correlation Coefficient (MCC) are calculated. Accuracy quantifies the overall proportion of correct predictions, while recall measures the model's ability to correctly identify positive instances. F1-score, a harmonic mean of precision and recall, offers a balanced evaluation of both metrics. Specificity assesses the model's capacity to correctly identify negative instances, and the MCC provides a correlation-based measure of the model's predictive power, considering both true and false positives and negatives. The equation of these metrics are as follows:

\begin{itemize}
    \item True Positive: Correct prediction of tumor tissue as tumor \\
    \item False Positive: Wrong prediction of ordinary tissue as tumor \\
    \item True Negative: Correct prediction of ordinary tissue as non-tumor \\
    \item False Negative: Wrong prediction of tumor tissue as non-tumor \\
    \item $Accuracy = (TN+TP)/(TP+FP+TN+FN)$ \\
    \item $Recall = TP/(TP+FN)$ \\
    \item $Specificity = TN/(TN+FP)$  \\
    \item $Precision = TP/(TP+FP)$ \\
    \item $F1$-$score = 2\times(Precision \times Recall)/(Precision + Recall)$\\
    \item \textit{Matthews Correlation Coefficient} $ = \frac{TP \times TN - FP \times FN}{\sqrt{(TP+FP)(TP+FN)(TN+FP)(TN+FN)}}$\\
\end{itemize}

\subsection{Experiment Settings}

For our experimental setup, a stratified 80/20 split of the training data is employed to create training and validation subsets, respectively. To address data imbalance, we take a five-fold cross-validation strategy. In each fold, one-fifth of the data is held out for validation, while the remaining data is used to train a model. The predictions from these five models are then averaged to create an ensemble. In the testing phase, we divide each fold based on specific indices to obtain varied proportions of data, rather than randomly selecting 20\% of the data for validation. We employ cross entropy loss as our loss function. To mitigate the impact of class imbalance during training, we assign the class weights to the loss function. To prevent overfitting and expedite training, an early stopping criterion is implemented, terminating the training process when validation loss ceases to improve after a maximum of 300 epochs. The Adam optimizer is selected for model optimization and initialized with a learning rate of $1e^{-4}$ and beta parameters of $(0.9, 0.999)$. To balance computational efficiency and model performance, a batch size of 64 is adopted. The PyTorch deep learning framework is served as the foundation for model development, with all experiments are executed on a single NVIDIA RTX 3090 GPU equipped with 24GB of video memory.

\subsection{Results}
Our model demonstrates promising performance on the local validation set, comprising 20\% of the training data, regarding evaluation metrics, with an accuracy, recall, and F1-score of 0.9872, a high specificity of 0.9974, and a strong Matthews Correlation Coefficient (MCC) of 0.9828. When evaluated on the more challenging Synapse online validation platform, the model achieves an accuracy of 0.392229, a recall of 0.392229, and a F1-score of 0.392229, indicating a inconsistent ability to correctly identify instances under the target condition. Notably, our model exhibits perfect specificity of 0.898704, demonstrating an exceptional capacity to correctly classify negative cases. Moreover, a MCC of 0.255267 is obtained, signifying a poor positive correlation between predicted and actual values and highlighting our model's overall predictive power.

\section{Conclusion}\label{conclusion}
Glioblastoma, characterized by its aggressive nature and intricate heterogeneity, presents a formidable challenge in accurate diagnosis and subsequent treatment selection. Traditional histopathological methods, while valuable, often fall short in comprehensively capturing the complex nature of this disease. This study explores the potential of deep learning as a tool to enhance glioblastoma diagnosis by participating in the BraTS-Path Challenge 2024. Our approach leverages a pre-trained model, followed by meticulous fine-tuning on the BraTS-Path training dataset. The resulting model demonstrates promising performance on the validation set, as assessed by the Synapse online platform. Our model achieves poor results on online validation set, including an accuracy, recall, and F1-score of 0.392229, perfect specificity of 0.898704, and an MCC of 0.255267, these results suggest limited generalization on online validation set. To enhance model reliability, future research should concentrate on developing models capable of consistent performance when applied to data from various sources.

\section*{Acknowledgements}
This work was supported by Ningbo Major Science \& Technology Project under Grant 2022Z126. (Corresponding: Ying Weng.)

\bibliographystyle{splncs04}
\bibliography{ref.bib}

\end{document}